\documentclass[journal=jacsat,manuscript=article]{achemso}
\usepackage{amssymb,amsmath}
\usepackage{graphicx}
\usepackage{color}
\usepackage{natbib}
\usepackage{hyperref}
\usepackage[utf8]{inputenc}

\usepackage[normalem]{ulem}

\usepackage[version=3]{mhchem}

\newcommand{\eps}{\varepsilon}      
\newcommand{\VB}[1]{\mathbf{#1}} 

\newcommand{\affilITMO}{Department of Physics and Engineering, ITMO University,  St.-Petersburg 197101, Russia}

\author{Ilya.~A.~Volkov}
\affiliation{\affilITMO}

\author{Roman~S.~Savelev}
\affiliation{\affilITMO}
\email{r.savelev@metalab.ifmo.ru}

\abbreviations{....}
\keywords{spin-photon interface, optical waveguides, chiral quantum optics, unidirectional emission}

\title[An \textsf{achemso} demo] {Unidirectional coupling of a quantum emitter to a subwavelength grating waveguide with engineered stationary inflection point}

\begin{document}

\begin{abstract}
In this work, we propose an approach for the design of a waveguide structure that allows for efficient and highly asymmetric coupling of the quantum sources with circularly polarized transition dipole moments to the guided mode of the structure. The approach is based on the mixing of the two quasi-degenerate modes of a periodic waveguide with an auxiliary single-mode waveguide leading to the formation of the dispersion with a stationary inflection point and consequently to the high coupling efficiency of this mode with a dipole source. We show that the distribution of the field polarization inside the waveguide is relatively homogeneous maintaining the circular polarization in a large area. Consequently, this leads to a high degree of tolerance of the coupling asymmetry and strength to the position of the quantum emitter. We believe, that our results will extend the variety of designs of the efficient chiral nanophotonic interfaces based on planar semiconductor nanostructures.
\end{abstract}

\section{Introduction}
Tailoring of the light-matter interaction at the nanoscale has attracted a lot of attention in the last decades due to increased technological capabilities of creating various optical nanostructures and their potential in the development of compact and scalable quantum integrated components based on the deterministic nanophotonic interfaces between the quantum emitters (QE) and light. One of the recently proposed ways to achieve this goal is based on the so-called chiral light-matter interaction~\cite{LodahlNature2017}, that manifests itself in an asymmetric interaction between the QEs with circularly polarized (CP) dipole transition moments and localized optical modes propagating in opposite directions.\par

All studies that utilize effects based on the chiral light-matter interaction can be categorized based either on the type of the QEs or on the type of the photonic nanostructures employed in the design. From the ``matter'' point of view one can distinguish a few main platforms including cold atoms~\cite{JungePRL2013,SayrinPRX2015,ScheucherScience2016}, semiconductor quantum dots~\cite{LodahlQuantum2018} and two-dimensional materials~\cite{ChenAdvOptMat2020}. Each of them, however, possesses specific features and limitations related to their experimental realization. Therefore, quite a variety of photonic nanostructures were theoretically and experimentally investigated in application to these platforms, including photonic crystal waveguides~\cite{YoungPRL2015, SollnerNatNanotech2015}, homogeneous optical nanowaveguides~\cite{LuxmooreAPL2013,ColesNatComm2016,HurstNL2018,JavadiNatNano2018,XiaoLPR2021}, whispering gallery-type resonators~\cite{JungePRL2013,ShomroniScience2014,Martin-CanoACSPhot2019} and more recently topological semiconductor waveguides and resonators~\cite{BarikPRB2020,MehrabadAPL2020,MehrabadOptica2020,RuanJLT2021}. Arguably one of the most suitable technologies for the practical purposes is based on a semiconductor GaAs platform that allows for integration of planar optical waveguides and cavities with quantum dots and efficient control of their emission properties~\cite{LodahlRMP2015,DietrichLPR2016,LodahlQuantum2018}. Several studies have experimentally demonstrated asymmetry of coupling of a quantum dot to a waveguide close to 100\%~\cite{FeberNatComm2015,SollnerNatNanotech2015,ColesNatComm2016,PregnolatoAPLPhot2020,XiaoLPR2021}. In addition, the strength of the coupling can also be increased when employing a cavity or slow-light modes of a photonic crystal (PhC) waveguides or cavities~\cite{FeberNatComm2015,SollnerNatNanotech2015,BarikPRB2020,MehrabadAPL2020,MehrabadOptica2020}. However, both coupling asymmetry and strength in such systems remain sensitive to the position of the QE due to the rather inhomogeneous field distribution in the unit cell of the PhC.

Lately, an alternative to the PhC-based systems in the form of various nanostructures composed of Mie resonant dielectric and semiconductor nanoparticles is being extensively explored~\cite{KrasnokSPIE2015,KivsharScience2016}. Such bottom-up approach allows to design structures with different geometry and with different optical properties by careful tuning of their individual building blocks and coupling between them. This includes single nanoantennas and their ensembles for control of the QE luminescence radiation patterns~\cite{RegmiNL2016,BouchetPRAppl2016,RutckaiaNL2017,ZaloginaNS2018,Sanz-PazNL2018}, optical cavities and waveguides composed of dielectric and semiconductor nanoparticles for the Purcell enhancement and lasing~\cite{KrasnokAPL2016,KrasnokPRAppl2018,HoangNL2020,RutckaiaACSPhot2020}, as well as active metasurfaces~\cite{StaudeACSPhot2019,VaskinNano2019}. In this context, periodic dielectric waveguides provide vast opportunities for dispersion engineering~\cite{HalirLPR2015, KuznetsovNL2017,ChebenNature2018} and achieving slow light regimes for efficient modulation of optical signals~\cite{DingACS2020} as well as polarization control due to the mode degeneracy engineering~\cite{YermakovIEEE2019,SavelevJAP2019,SavelevPRB2020}.

In this work, we propose an approach for the design of an optical nanostructure based on a periodic waveguide that can serve as an efficient chiral interface between the propagating waveguide modes and the circularly polarized dipole sources embedded in the structure. This approach relies on two factors: the presence of two quasi-degenerate orthogonal modes in a periodic waveguide, and a specific type of dispersion of the whole structure possessing a stationary inflection point (SIP). The combination of these factors allows increasing the strength of the light-matter interaction while at the same time keep the interaction highly asymmetric. Moreover, unlike other designs studied before, the coupling of a quantum source to the mode of the proposed structure is substantially more tolerant to the position of the source, which makes it beneficial from the practical point of view. We believe, that the presented results will allow for the development of the advantageous bottom-up designs of compact chiral nanophotonics interfaces for efficient and scalable integrated quantum circuitry.

\section{Formulation of the idea}

\begin{figure}[t]
    \includegraphics[width=0.5\columnwidth]{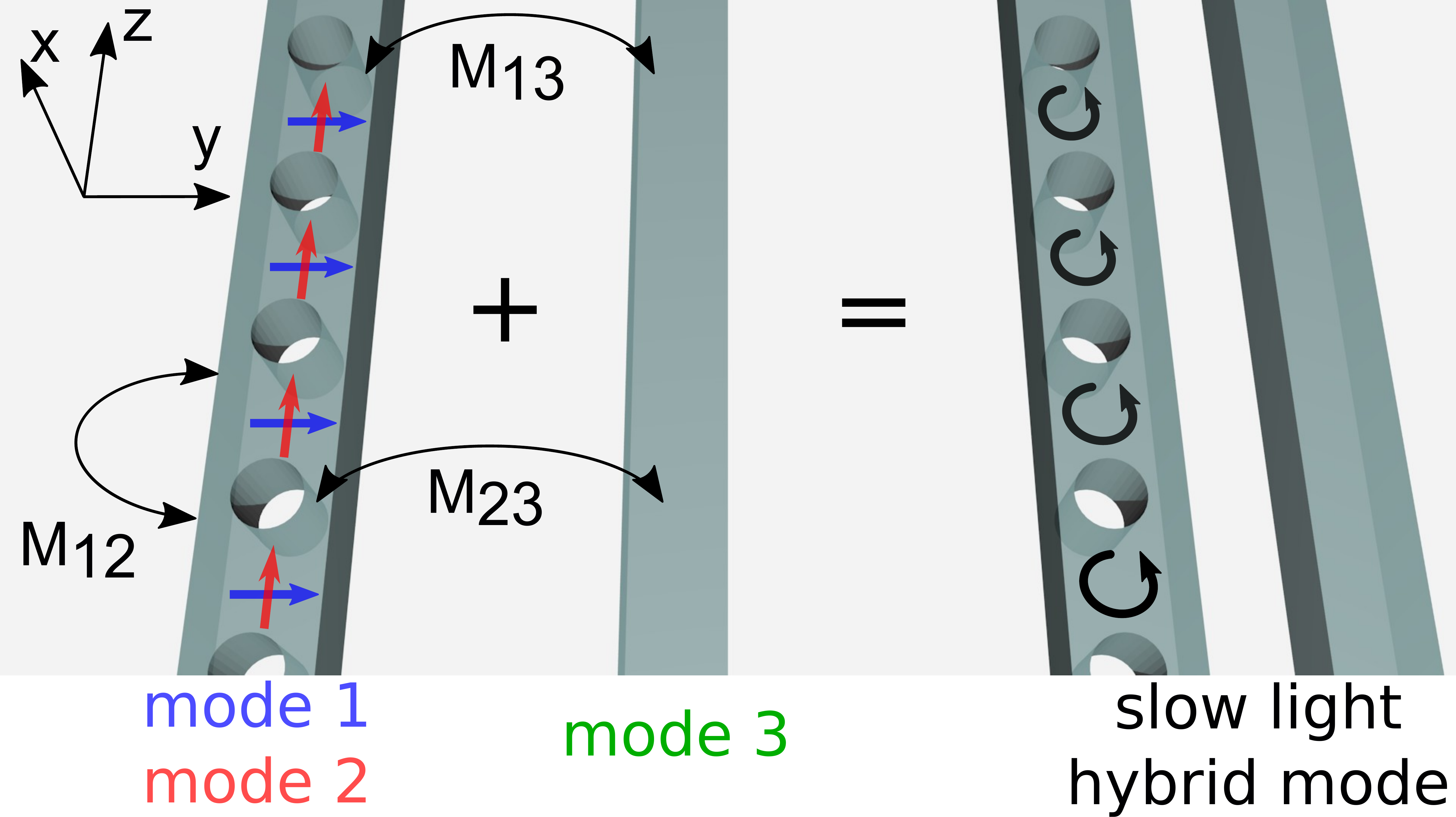}
    \caption{(a) Scheme of the considered system: two orthogonal modes of a periodic waveguide and a single mode of a homogeneous waveguide couple with each other forming a single supermode with low group velocity and local circular polarization of the field inside the periodic waveguide. Arrows indicate the dominant polarization of the field.} 
    \label{fig:scheme}
\end{figure}

The design of the proposed waveguiding structure that allows for efficient and uniform coupling with circularly polarized dipole emitters is based on a specific type of the waveguide dispersion $\omega(k)$ that contains a stationary inflection point, i.e. a point at which the group velocity $d\omega/dk$ and group velocity dispersion $d^2\omega/dk^2$ vanish. It is known that such type of dispersion emerges when at least three modes are coupled in a specific way forming an exceptional point of degeneracy of the third order~\cite{GutmanOE2012,NadaPRB2017,NadaArxiv2020}. In the current work we consider a structure in which two of the three modes belong to the same periodic waveguide and the third one belongs to the auxiliary spatially separated homogeneous waveguide, as schematically shown in Fig.~\ref{fig:scheme}. If two quasi-degenerate modes of the periodic waveguide have local electric field polarization orthogonal to each other, then the polarization of the mixed mode of the coupled waveguides will be elliptical in general, with the specific type of polarization depending on the type of the coupling between the modes. Further, if the characteristics of the third mode are chosen in a proper way, the mixed mode will possess a SIP in the dispersion.

Such a mode of the structure exhibits several features beneficial for the chiral light-matter interaction. First, the group velocity of the mode tends to zero at a certain frequency, providing a potential for an arbitrarily strong coupling of the point dipole source to the guided structure. Second, by tuning the geometrical parameters of the structure, one can effectively control a type of the coupling between the modes, which determines the local field polarization of the mode. Thereby, one can design a waveguide with circular local polarization of the electric field and thus reach the fully asymmetric coupling of the dipole source to the modes propagating in forward and backward direction along the waveguide. Finally, the distribution of the field of uncoupled modes, which can be considered as rather homogeneous (e.g. compared to the exponentially decaying waves) are inherited by the supermode. This allows to maintain efficient and asymmetric coupling to a single quantum emitter located not only at a specific point, but rather in a random point in the relatively large area.

\section{Engineering of the dispersion and polarization in coupled waveguides}

\begin{figure}[t]
    \includegraphics[width=1\columnwidth]{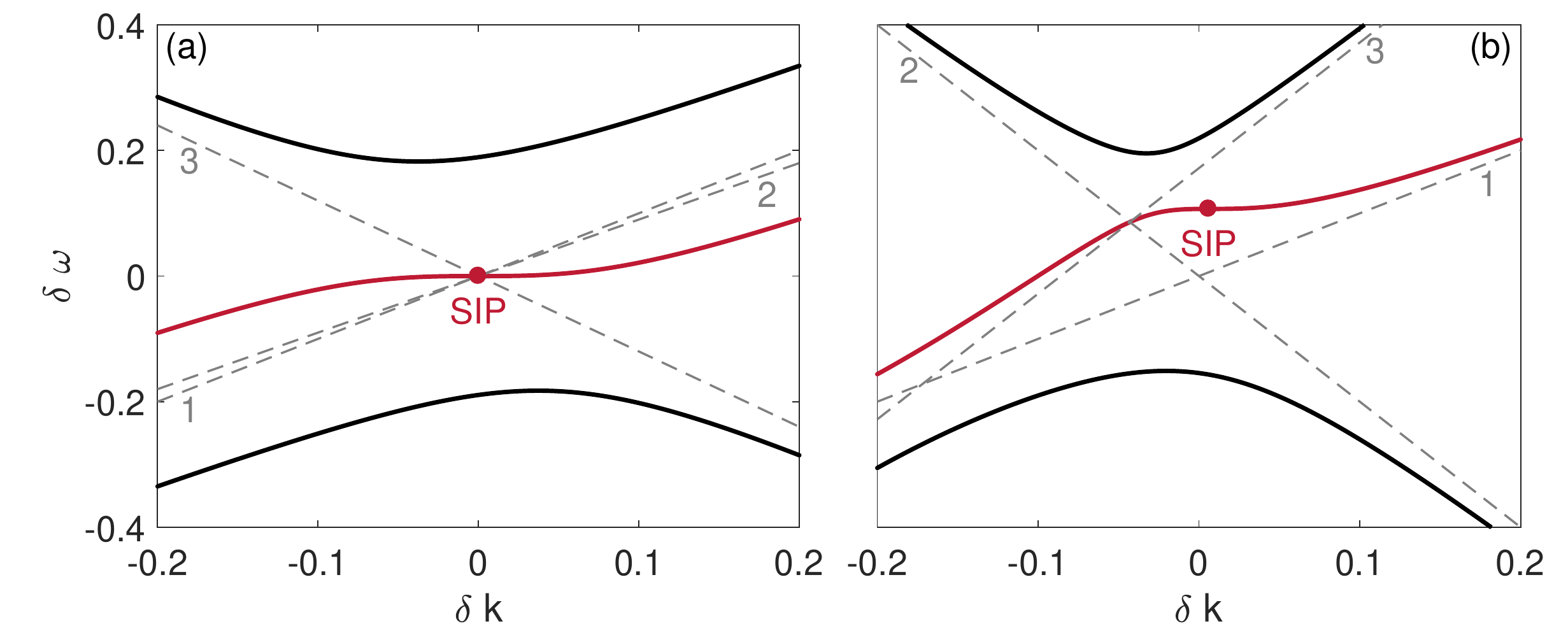}
    \caption{Dispersion of the modes of coupled waveguides (solid curves) calculated within the framework of the CMT. Stationary inflection point is marked with a red circle. Dashed lines correspond to the modes of the isolated waveguides. Parameters of the system: (a) group velocities $v_{g1}=1$, $v_{g2}=0.9$, $v_{g3}=-1.2$, frequency detuning of the third mode $\Delta_3=0$, coupling constants $M_{12}=-0.1258$, $M_{13}=0.1i$, $M_{23}=0.1$. The eigenvector at the SIP is $(1,i,1.258)^T$. (b) $v_{g1}=1$, $v_{g2}=-2$, $v_{g3}=2$, $\Delta_3=0.172$, coupling constants $M_{12}=-0.145i$, $M_{13}=0.077$, $M_{23}=0$. The eigenvector at the SIP is $(1,1.2i,-0.97)^T$. } 
    \label{fig:Disp_CMT}
\end{figure}

The group velocity of the coupled mode is determined by the sum of the group velocities of the isolated modes, therefore it can be equal to zero only if at least one of the modes is a backward wave, i.e. has a negative group velocity. This can be realized in two different ways. First, we consider the case when two quasi-degenerate forward waves belong to a periodic waveguide, while the backward wave is supported by a homogeneous waveguide, as shown in Fig.~\ref{fig:Disp_CMT}(a) with dashed lines. From what follows, we assume that indices ``1'',``2'' correspond to the modes of a periodic waveguide, and the index ``3'' corresponds to a homogeneous waveguide. Although a homogeneous waveguide generally supports modes only with positive group velocity, in the considered system the mode effectively becomes a backward wave after the dispersion folding, that occurs due to its coupling with the periodic waveguide (see Supporting Information, Sec. A). In an alternative design, the mode with a negative group velocity belongs to a periodic waveguide, while the mode of a homogeneous waveguide is a forward one, Fig.~\ref{fig:Disp_CMT}(b).

Although these two designs have qualitative differences that are explained further, the physical insight into important determinants of the interaction between the waveguide modes can be gained from the analysis carried out within the same framework of the conventional coupled-mode theory. Once the characteristics of the modes of the isolated waveguides are known, e.g. from numerical simulations, one can find the eigenfrequencies $\omega$ of the coupled supermodes from the matrix equation (see Supporting Information, Sec. B for the details):
\begin{equation}
\VB{M}\VB{A} =\omega \VB{A},
\label{CMT}
\end{equation}
where $\VB{M}$ is the 3$\times$3 coupling matrix, $\VB{A}$ is the vector of the amplitudes $a$ of the three interacting modes $\VB{A} = (a_1,a_2,a_3)^T$. Generally, all variables in Eq.~\eqref{CMT} depend on the Bloch wavenumber $k$, thus allowing to obtain a dispersion $\omega(k)$ of the coupled modes and their field distributions. The elements of the coupling matrix $\VB{M}$ can be adjusted in a wide range of values by tuning the geometrical parameters of the considered structures. For instance, the diagonal elements are mostly determined by the dispersion of the modes of the isolated waveguides, and they can be tuned by changing the geometrical parameters of the isolated waveguides. The non-diagonal element $M_{12}$ describes the coupling between the modes of the periodic waveguide and its value is mostly influenced by the type of the asymmetry (with respect to $xz$ plane) introduced in the initially symmetric periodic waveguide. In turn, the elements $M_{13},M_{23}$ that describe the interaction between the modes of periodic and homogeneous waveguides are mostly determined by the distance between the waveguides. The relations between the geometrical parameters of the system and the elements of the matrix $\VB{M}$ are described in Supporting Information, Sec. C.

Using the formalism of the coupled mode theory one can analyze the requirements to the elements of the coupling matrix that allow to achieve the desired characteristics of the waveguide, i.e. dispersion with a SIP and circular polarization of the local electric field in the periodic waveguide. To derive such requirements, we impose the following conditions on the eigenmode of the Eq.~\eqref{CMT} at the wavenumber $k_0$:
\begin{equation}
    \dfrac{d\omega}{dk}\Bigr|_{k=k_0}=0,\;\;\dfrac{d^2\omega}{dk^2}\Bigr|_{k=k_0}=0,
    \label{eq:SIP_condition}
\end{equation}
\begin{equation}
    \VB{A}(k_0) = (1,i\alpha,\beta)^T,
    \label{eq:CP_condition}
\end{equation}
where $\alpha$ and $\beta$ are arbitrary real numbers (the phase of the third mode can be chosen at will). The Eq.~\eqref{eq:SIP_condition} specifies the existence of the SIP, $\omega(k) \approx \omega(k_0) + \gamma(\delta k)^3$, at the point $k_0$, while the eigenvector~\eqref{eq:CP_condition} ensures the desired polarization of the field in the periodic waveguide. Note that we are interested in the circular polarized electric field rather than circular polarized eigenvector $\VB{A}$ of the coupled mode, therefore the coefficient $\alpha$ in a realistic design should be imaginary but its amplitude should not necessarily be equal to $1$.

In Fig.~\ref{fig:Disp_CMT}(a,b) we show two dispersion diagrams of the coupled waveguides with the parameters that satisfy the conditions~(\ref{eq:SIP_condition}-\ref{eq:CP_condition}). In the panel (a), the third mode has negative group velocity, while in the panel (b) -- the second one. In the first case, we have found that the conditions~\eqref{eq:SIP_condition},\eqref{eq:CP_condition} can be fulfilled when, first, all three modes intersect at the same point $k_0=0$, and second, the elements $M_{12}$ and $M_{23}$ are real, while $M_{13}$ is imaginary, resulting in the eigenvector with $\alpha = \mathrm{Im}(M_{13})/M_{23}$ and $\beta=-M_{12}/M_{23}$. In this case, the $\pm \pi/2$ phase difference between the modes ``1'' and ``2'' is ensured by the zero phase of the coefficient $M_{12}$, while the magnitude of $\alpha$ is defined by the ratio $|M_{13}/M_{23}|$, which is of the order of 1 in realistic systems due to the quasi-degeneracy of the first two modes. In the second case, Fig.~\ref{fig:Disp_CMT}(b), the mode ``2'' has negative group velocity. In this case it typically interacts weakly with the mode ``3'', therefore we assume that $M_{23}=0$. In that case, first, we need to properly tune the frequency detuning of the third mode $\Delta$ and the amplitudes of the coupling constants $M_{12}$, $M_{13}$, and second, in contrast to the first design the $M_{12}$ constant should be imaginary, see details in the Supporting Information, Sec. B.

Although, in these simple calculations we assumed that non-diagonal elements of the matrix $\VB{M}$ do not depend on $k$, and diagonal elements are linear functions of $k$, such simplification allowed us to identify the critical requirements to the waveguide design used in the following section.

\section{Designs of the waveguides based on structured semiconductor membranes}

\begin{figure*}[t]
    \includegraphics[width=1\textwidth]{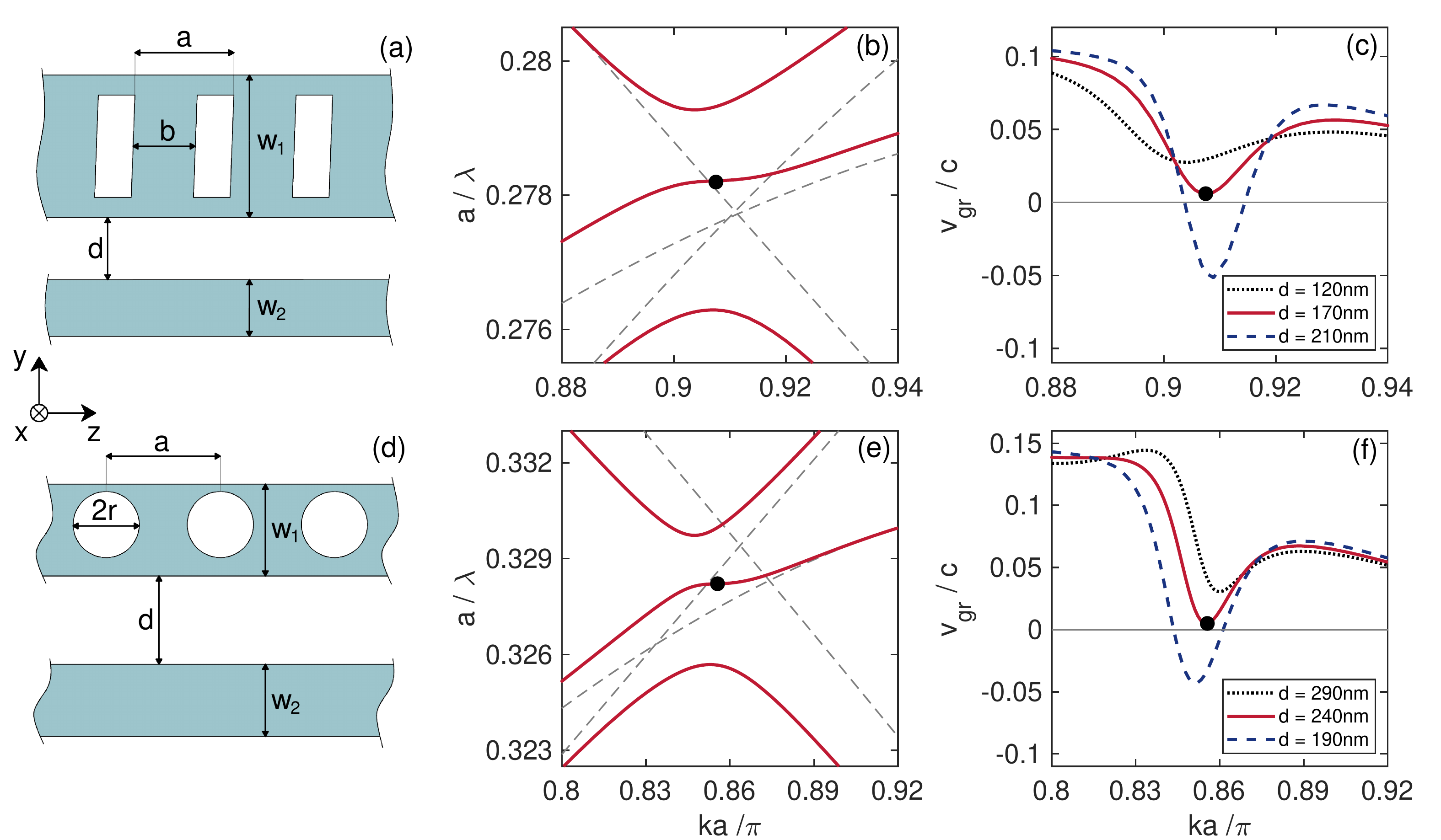}
    \caption{(a,d) Schemes of the guided structures under consideration (top view); (a) $a=270$~nm, $b=170$~nm, $w_1=390$~nm, $w_2=154$~nm, $d=170$~nm, thickness of the membrane $h=220$~nm, shift between the upper and lower vertices of the hole in $z$ direction is 10~nm; (b) $a=310$~nm, $r=90$~nm, $w_1=250$~nm, $w_2=197$~nm, $d=240$~nm, $h=245$~nm, shift of the holes in $y$ direction $\delta y= 15$~nm. (b,e) Dispersion of the modes of the isolated (gray dashed curves) and coupled (solid red curves) waveguides for the structures shown in (a,d), respectively. Black dots indicate the SIPs. (c,f) Group velocity of the mode with a SIP as a function of the wavenumber for 3 values of the gap between the waveguides.}
    \label{fig:dispersions}
\end{figure*}
In order to demonstrate the feasibility of the proposed approach in the design of realistic optical structures we have developed two designs of the coupled waveguides based on the nanostructured GaAs membranes with refractive index $n=3.5$. The parameters of the structures were tuned to a wavelength around $950$~nm corresponding to the emission wavelength of the InGaAs quantum dots. The schemes of the structures are shown in Fig.~\ref{fig:dispersions}(a,d) with the geometrical parameters given in the caption. In Fig.~\ref{fig:dispersions}(b,e) we show the dispersion properties of isolated and coupled waveguides with dashed gray and thick solid red curves, respectively. As it follows from the coupled mode theory, the distance between the waveguides $d$ mainly affects the relative contribution of the mode of the homogeneous waveguide to the supermode, which in turn affects its group velocity. This can be observed in Fig.~\ref{fig:dispersions}(c,f): by adjusting the distance $d$, it is formally possible to achieve an ideal SIP. We have chosen the optimal values of $d$ so that the structure supports a single mode with the group velocity value around $c/200$ (where $c$ is the speed of light) near the operational frequency marked with black circles in Figs.~\ref{fig:dispersions}(c,f).

Besides the evanescent coupling between the waveguides the modes of the periodic waveguide were also coupled with each other by introducing a slight asymmetry to the waveguide: in the first design the rectangular holes were modified by the slight shear deformation in $z$ direction, and in the second design the circular hole were shifted in $x$ direction by a small distance. By doing this the coupling constant $M_{12}$ satisfied the previously derived conditions and the modes of the periodic waveguide were coupled with the $\pi/2$ phase difference, providing the local circular polarization of the electric field. In order to quantify the resulting polarization properties of the waveguide and slow-light enhancement of a dipole emission at the same time, we have performed calculations of the coupling strength of a dipole QE to a waveguide mode $\gamma_{WG}$ using Fermi's golden rule, i.e. assuming weak interaction. Within this framework, $\gamma_{WG}$ normalized by the decay rate in the bulk material $\gamma_0$ is found using the following formula~\cite{MangaRaoPRB2007,FeberNatComm2015}:
\begin{equation}
    \dfrac{\gamma_{WG}^{\pm}}{\gamma_0} = \dfrac{3\pi c^3a|\VB{E}^{\pm}(\VB{r}_d)\cdot \VB{d}^*|^2}{2\omega_d^2\sqrt{\eps}v_g},
    \label{Fermi}
\end{equation}
where $c$ is speed of light, $a$ is the waveguide period, $\omega_d$ is the dipole oscillation frequency, $v_g$ is the eigenmode group velocity, $\eps$ is the permittivity of the waveguide material, $\VB{r}_d$ is the radius vector that defines the position of the dipole with unit vector $\VB{d}$, and $\VB{E}^{\pm}$ is the electric field distribution of the modes propagating in positive and negative $z$ direction, respectively, normalized in such a way that $\int_{\text{unit cell}} \eps |\VB{E}|^2 dV = 1$. We calculate this quantity for the fixed polarization of the dipole source $\VB{d}=(0,1,i)/\sqrt{2}$ located in the plane $x=\pm 90$~nm and oscillating at the frequency that corresponds to the SIP.

\begin{figure*}[t]
    \includegraphics[width=1\textwidth]{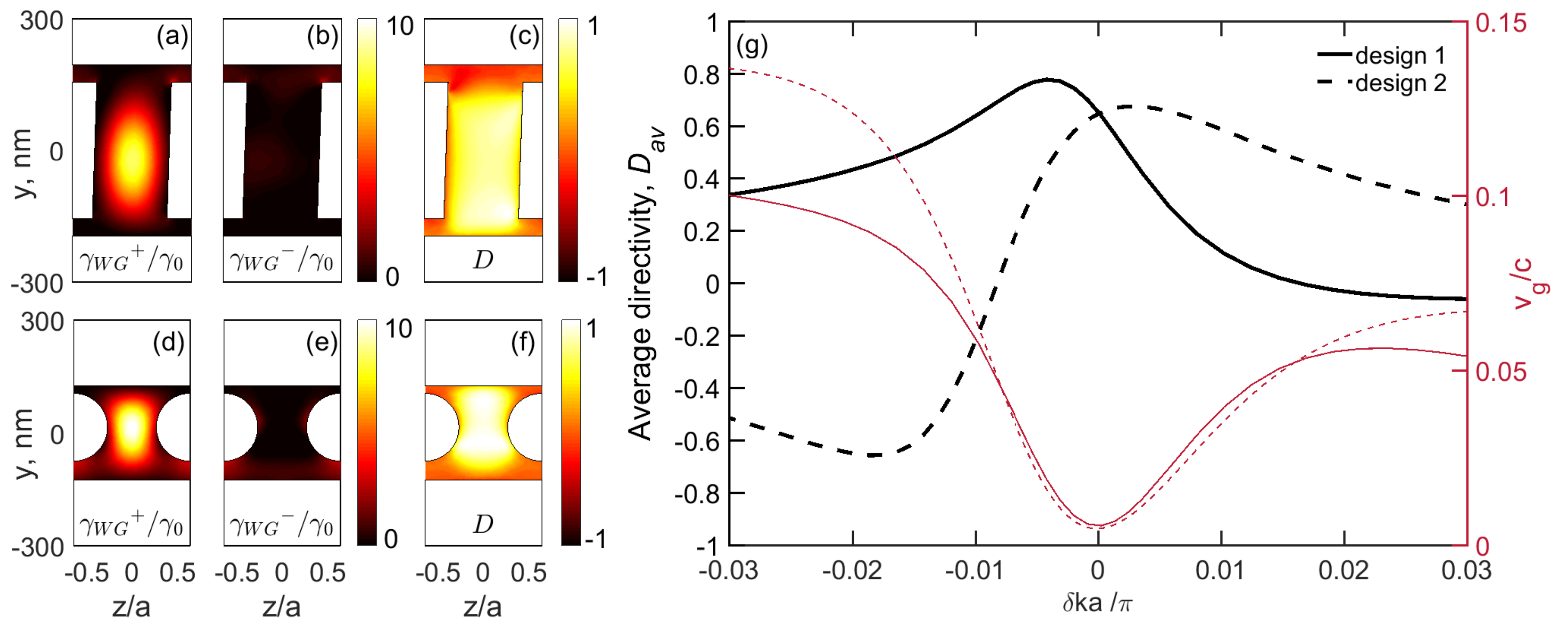}
    \caption{(a,b,d,e) Emission rate into the slow-light mode propagating in (a,d) positive $z$ and (b,e) negative $z$ direction as a function of the position of a dipole source in the cross section $x=90$\;nm of the periodic waveguide. (c,f) Directivity of the emission. (a-c) correspond to the first design, (d-f) -- to the second one.  (g) Average directivity $D_{av}$ (black curves, left axis) and normalized group velocity (red curves, right axis) as a function of the normalized wavenumber detuning. Zero detuning corresponds to the minimum group velocity. Solid curves correspond to the first design, dashed curves -- to the second one.}
    \label{fig:chirality}
\end{figure*}

The results of calculations are presented in Fig.~\ref{fig:chirality}. One can observe that the coupling rate of the dipole source to the mode propagating in positive $z$ direction $\gamma_{WG}^+$ normalized by the rate of emission in bulk semiconductor $\gamma_{0}$ [panels (a,d)] reaches values up to 10 in both designs, and it is much stronger than the coupling to the mode propagating in negative $z$ direction, $\gamma_{WG}^-$ [panels (b,e)], in most parts of the periodic waveguide cross-section. The directivity of the emission $D$, calculated as $D = (\gamma_{WG}^+ - \gamma_{WG}^-)/(\gamma_{WG}^+ + \gamma_{WG}^-)$ reaches values close to 1 in the most part of the waveguide cross section. In order to better illustrate the robustness of the asymmetry to the position of the QE, we have introduced the figure of merit that characterizes the average directivity of emission of a randomly located point dipole source defined as $D_{av}=\dfrac{\langle \gamma_{WG}^+ - \gamma_{WG}^- \rangle}{\langle \gamma_{WG}^+ + \gamma_{WG}^- \rangle}$, where averaging is performed over the cross-sectional area of the periodic waveguide in the plane $x=90$~nm. Fig.~\ref{fig:chirality}(g) shows that for the chosen parameters $D_{av}$ reaches $\approx 0.65$ in both cases. Such high value of the average asymmetry becomes possible due to the locked $\pi/2$ phase difference between two guided modes with orthogonal polarization of the electric field, achieved by proper engineering of the coupling between the waveguides.

\section{Estimation of the $\beta$ factor}

Although the main characteristics of the coupling between the dipole source and the waveguide can be well understood from the eigenmode simulations, these characteristics lack the information about the coupling strength of the dipole to all other modes of the surrounding enviroment. In order to estimate the $\beta$-factor, which is the ratio of the decay rate to the given waveguide mode and the total decay rate $\beta=\gamma_{WG}/\gamma$, we perform numerical simulations of the system with the quasi-point dipole-like source. Since there is no qualitative difference in the characteristics of both designs, further we present the results only for the second design.

Calculation of the $\beta$-factor to the infinite periodic waveguide is not a straight-forward problem due to necessity of applying of the special boundary conditions that will absorb the propagating mode~\cite{JavadiJOSAb2018}. Such problem is especially complicated when the considered guided mode has a slow group velocity~\cite{OskooiOE2008}. In our simulations we have considered a finite size waveguide with the adiabatic absorbers at the end of the waveguide fed by a circularly polarized dipole source with two possible polarizations $\VB{d}^{\pm}=(\VB{e}_y \pm i\VB{e}_z)/\sqrt{2}$ placed at the point with coordinates $\VB{r}=(90,-25,0)$~nm, where maximum coupling asymmetry was expected. From the simulations we have extracted the power emitted into waveguide modes propagating in positive and negative $z$ directions as well as the power radiated into free space and other modes of the system. In order to check the validity of the results we have performed several simulations varying the size of the waveguide and the absorbing part, making sure that the corresponding variation of the power coupled to the waveguide and $\beta$-factor becomes insignificant. Additionally, we have calculated the power emitted by the same source in the bulk semiconductor, $\gamma_0$, and compared the results of the $\gamma_{WG}/\gamma_0$ obtained in the simulations with the source and the eigenmodes simulations.

\begin{figure}[t]
    \includegraphics[width=0.75\columnwidth]{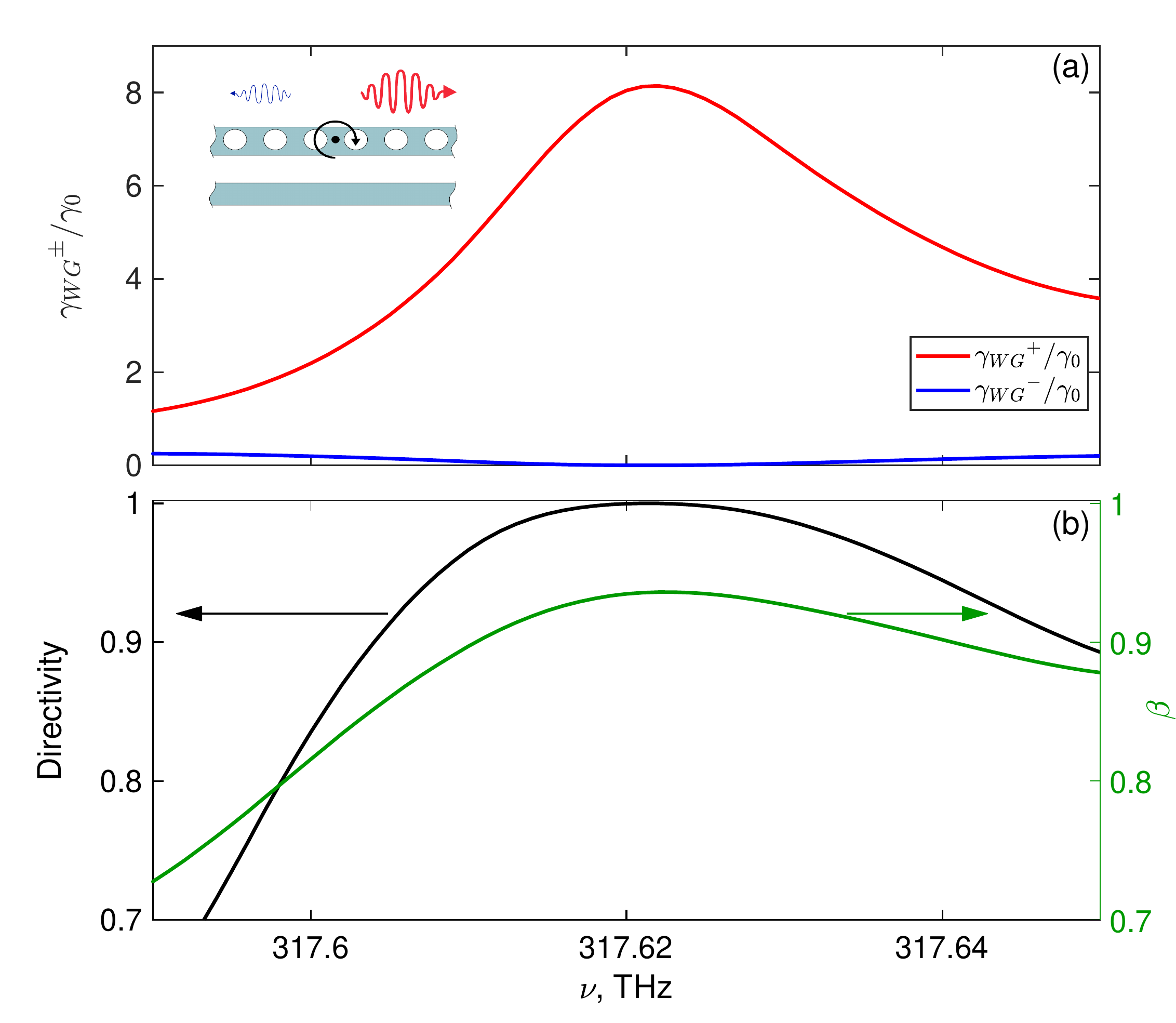}
    \caption{(a) Emission rate of a circularly polarized dipole source into the slow-light mode propagating in positive $z$ direction $\gamma_{WG}^{+}$ (solid red) and negative $z$ direction $\gamma_{WG}^{-}$ (solid blue) normalized by the emission rate in a bulk semiconductor $\gamma_0$ as a function of frequency. (b) directivity (black curve, left axis) and $\beta$-factor (green curve, right axis) as a function of frequency.} 
    \label{fig:source}
\end{figure}

The results of calculations of the emission enhancement factor into the waveguide modes $\gamma_{WG}^{\pm}/\gamma_0$ are presented in Fig.~\ref{fig:source}(a). The peak value of the $\gamma_{WG}^{+}/\gamma_0 \approx 8$ is in a good agreement with the $\gamma_{WG}^{+}/\gamma_0 \approx 8.5$ obtained from the eigenmode simulations for the given position and polarization of the source, Fig.~\ref{fig:chirality}(c), which confirms that the simulations of the system with the dipole source provide reliable characteristics of the structure. The substantial difference between $\gamma_{WG}^{+}$ and $\gamma_{WG}^{-}$ rates leads to the almost perfect directivity near the frequency corresponding to the SIP, see black curve in Fig.~\ref{fig:source}(b). The beta factor at the same time reaches values up to 94\%. Note that theoretically it is possible to increase the coupling to the waveguide modes, and consequently the $\beta$-factor, by fine tuning of the system parameters in an unlimited way. The current values were obtained for the experimentally achievable group indices as large as $200$.

\section{Summary}

To summarize we have demonstrated a possibility of enhancement of the chiral light-matter interaction in the periodic dielectric waveguides through simultaneous engineering of the stationary inflection point in the dispersion and tailoring local polarization of electric field by exploiting multi-mode regime of the waveguides. Highly efficient and directive emission of the circularly polarized dipole sources into the desired guided mode was demonstrated in numerical simulations. Along with the robustness of the system to the position of the source, this makes the proposed design promising for the further development of the planar integrated spin-photon interfaces enabled by the spin-orbit interactions of light.

\begin{suppinfo}
Details on the optical properties of the isolated waveguides; formulation of the coupled mode theory; relation between the geometrical parameters of the structure and the CMT.
\end{suppinfo}

\begin{acknowledgement}
This work was supported by the Russian Science Foundation, Grant No.~19-72-10129. R.S. acknowledges the support by the Grant from the President of Russian Federation MK-4418.2021.1.2.
\end{acknowledgement}

\bibliography{References}

\end{document}


\title{Unidirectional coupling of a quantum emitter to a subwavelength grating waveguide with engineered stationary inflection point: Supporting Information}

\newcommand{\affilITMO}{Department of Physics and Engineering, ITMO University,  St.-Petersburg 197101, Russia}

\author{Ilya.~A.~Volkov}
\affiliation{\affilITMO}

\author{Roman~S.~Savelev}
\affiliation{\affilITMO}
\email{r.savelev@metalab.ifmo.ru}


\maketitle
\tableofcontents

\section{A. Properties of the isolated waveguides}\label{sec:AppA}

The structures shown schematically in Figs.~3(a,d) in the main text are based on periodic waveguides that support two quasi-degenerate modes, coupled with a waveguide with a single mode in the spectral range of interest. The possibility to achieve a dispersion crossing of two modes with orthogonal dominant polarizations was demonstrated previously in the Refs.~\cite{SavelevJAP2019,SavelevPRB2020}. Such modes can be considered as two constituent elements that can be combined in an arbitrary way allowing for controllable tailoring of local polarization of the electric field in the plane parallel to the substrate.

\begin{figure*}[t]
    \includegraphics[width=1\textwidth]{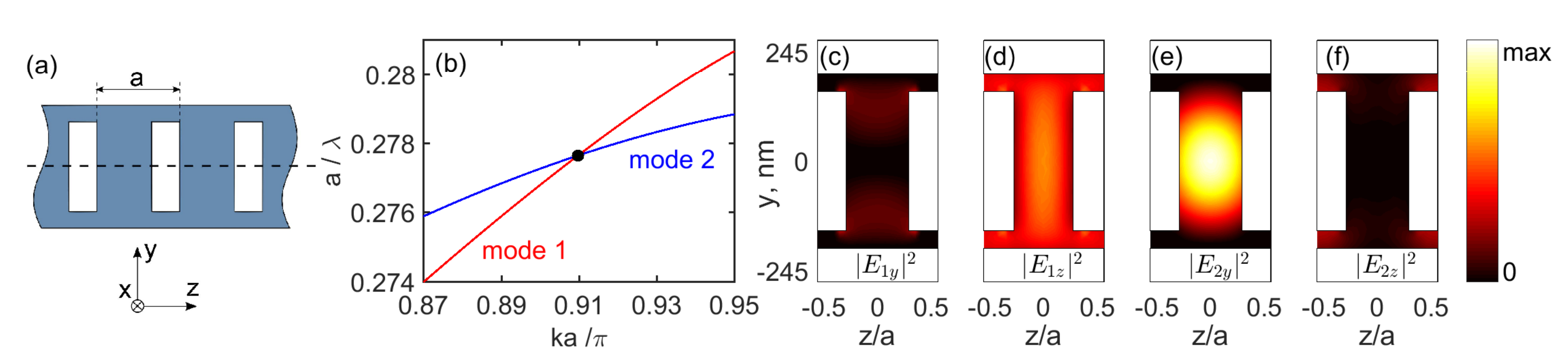}
    \caption{(a) Dispersion of a periodic waveguide with two intersecting modes with positive group velocities (top view), $\lambda$ is the wavelength, $k$ is the Bloch wavenumber. Thickness of the waveguide membrane $h=220$~nm, period is $a=270$~nm, dimensions of the rectangular hole are $100 \times 310$~nm, thickness of the waveguide $w_1=390$~nm. (c-f) Intensities of $y$ (c,e) and $z$ (d,f) electric field in the cross section of the unit cell in the plane $x=90$~nm for the (c,d) first modes, (e,f) second mode.} 
    \label{fig:APP_modes1}
\end{figure*}

In the first design we consider a periodic waveguide with two forward quasi-degenerate modes. Their dispersion is shown with red and blue curves in Fig.~\ref{fig:APP_modes1}. The dispersion of the modes can be approximated as following:
\begin{equation}
\begin{aligned}
& \omega_1\;\text{[rad/s]} \approx 2\pi \cdot 308.5\cdot10^{12} + 5.1\cdot10^7 \delta k - 21 \delta k^2,\\
& \omega_2\;\text{[rad/s]} \approx 2\pi \cdot 308.5\cdot10^{12} + 2.3\cdot10^7 \delta k - 19 \delta k^2,
\end{aligned}
\end{equation}
where $\delta k$ is the detuning of the wavvenumber from the crossing point.

Both of these modes possess two planes of symmetry: electric (tangential components of electric field is zero) plane of symmetry $x=0$, and electric (magnetic) plane of symmetry $y=0$ for the first (second) mode. Consequently, in the plane $x=0$ electric field in both modes has only $x$ component. However, when shifted by any distance in $x$ direction, one mode acquires (dominant) $y$ component of electric field, while another one $z$ component. Distributions of the $y$ and $z$ components of electric field at the distance $x=90$~nm are shown for both modes in Fig.~\ref{fig:APP_modes1}. 

\begin{figure*}[t]
    \includegraphics[width=0.6\textwidth]{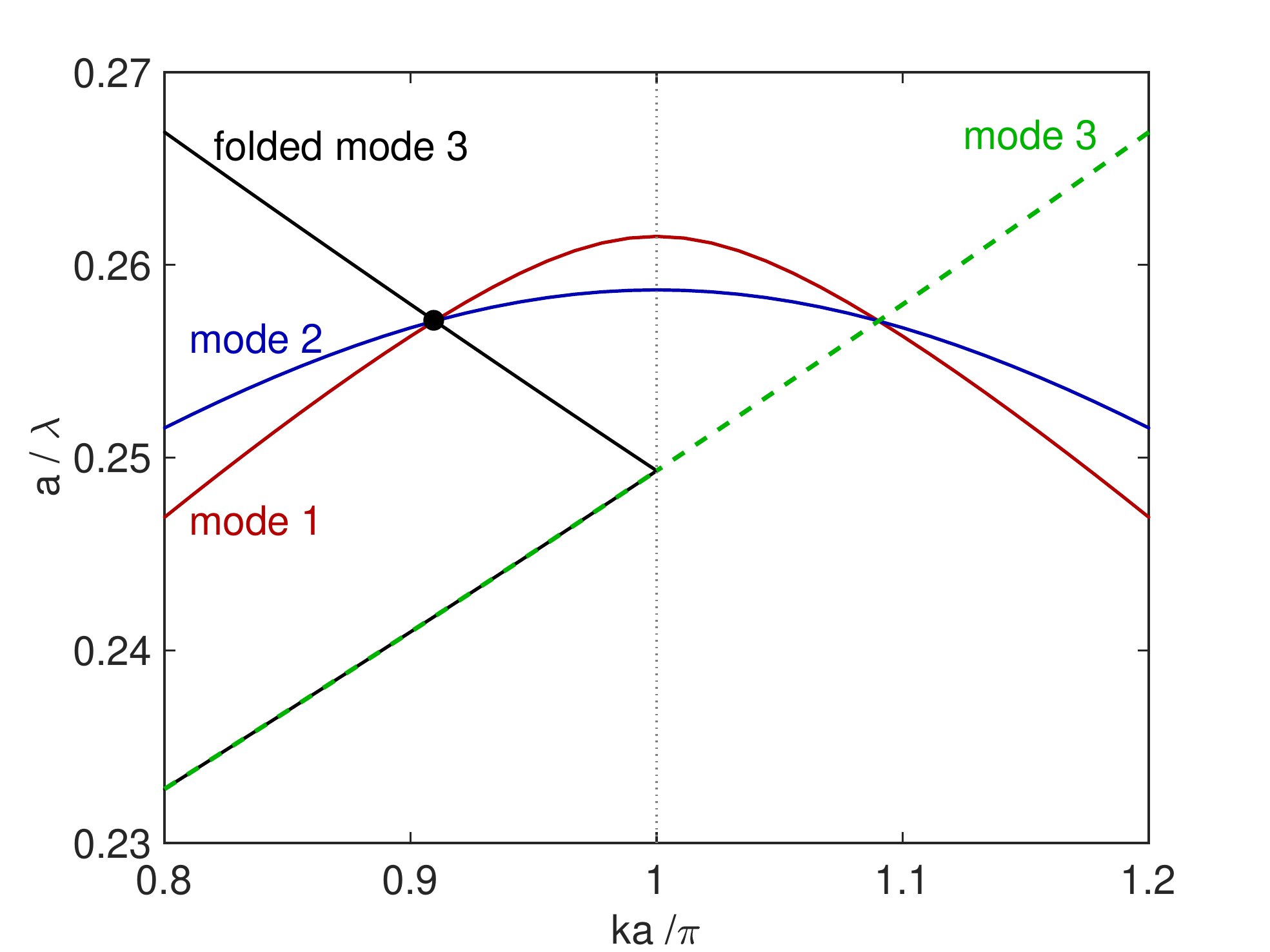}
    \caption{Dispersion of a periodic waveguide with two intersecting modes ``1'' and ``2'' with positive group velocities (red and blue curves) and a dispersion of a side-coupled strip waveguide (dashed green curve) with mode ``3''. Due to the dispersion folding, dispersion of the third mode reflects from the edge of the Brillouin zone ($ka/\pi=1$), as shown with the solid black curve, changes the sign of the group velocity and intersects with the first and second modes.}
    \label{fig:folding}
\end{figure*}

In order to achieve a zero group velocity of the supermode of the coupled waveguides, one has to compensate the positive group velocities of these modes by another mode with negative group velocity. To this end we employ a homogeneous strip waveguide with a single guided mode in the operational frequency range. Although dispersion $\omega(k)$ of a homogeneous waveguide is a monotonic function with a positive group velocity, introducing arbitrarily small periodicity $a$ in the direction of the waveguide axis $z$ makes the dispersion $\omega(k)$ a periodic function with the period equal to reciprocal lattice vector $2\pi/a$. Consequently, due to dispersion folding, i.e. reflection of dispersion branch from the edge of the Brillouin zone, some parts of the dispersion curves become effectively backward, as illustrated in Fig.~\ref{fig:folding}. In other words, there is a contradirectional coupling between the mode of the strip waveguide that propagates in the negative $z$ direction and the modes of the periodic waveguide that propagate in the positive $z$ direction. By tuning the width of the strip waveguide, we were able to achieve the crossing of the dispersion of a strip waveguide and the intersection point of the modes of the periodic waveguide, as shown in Fig.~\ref{fig:folding}. The interaction between these modes is further described within the framework of the coupled mode theory, see Sec. B.

\begin{figure*}[t]
    \includegraphics[width=1\textwidth]{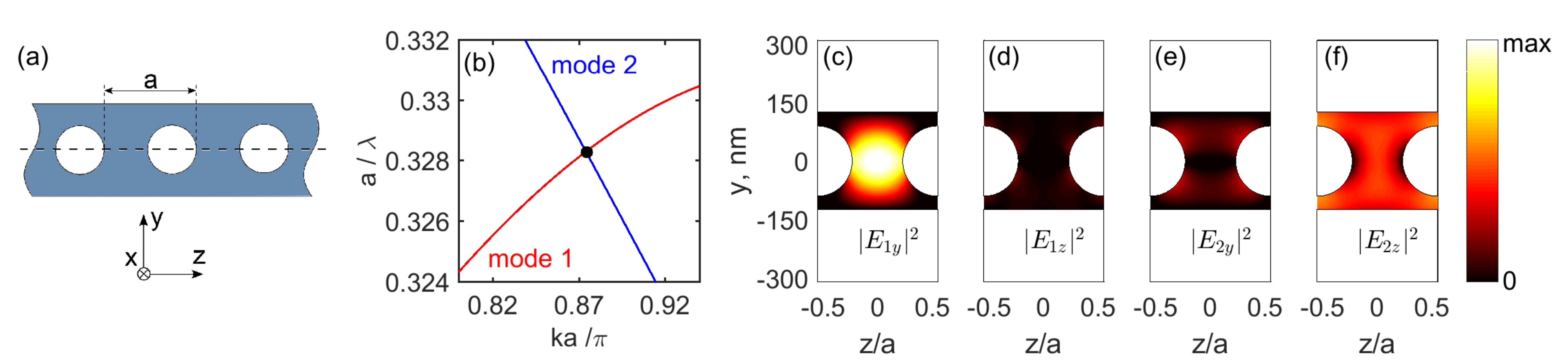}
    \caption{(a) Dispersion of a periodic waveguide with two intersecting modes with group velocities of opposite sign (top view), $\lambda$ is the wavelength, $k$ is the Bloch wavenumber. Thickness of the waveguide membrane $h=245$~nm, period is $a=310$~nm, radius of the holes is $90$~nm, thickness of the waveguide $w_1=250$~nm. (c-f) Intensities of $y$ (c,e) and $z$ (d,f) components of electric field in the cross section of the unit cell in the plane $x=90$~nm for the (c,d) first mode, (e,f) second mode.} 
    \label{fig:APP_modes2}
\end{figure*}

In the second design the modes of the periodic waveguide have different signs of the group velocities. The scheme of the waveguide, its dispersion and the field distributions at the crossing point are shown in Fig.~\ref{fig:APP_modes2}. The dispersions of the modes can be approximated as following:
\begin{equation}
\begin{aligned}
& \omega_1\;\text{[rad/s]} \approx 2\pi \cdot 317.7\cdot10^{12} - 5.51\cdot10^7 \delta k - 2.3 \delta k^2,\\
& \omega_2\;\text{[rad/s]} \approx 2\pi \cdot 317.7\cdot10^{12} + 2.26\cdot10^7 \delta k - 13.5 \delta k^2.
\end{aligned}
\end{equation}Again, in the plane $x=90$~nm (parallel to substrate), one of the modes has dominant $y$ component of the electric field, while the second one -- $z$ component.

\section*{B. Coupled mode theory}\label{sec:AppB}

\subsection{General formulation}

To analyze the relation between the geometrical parameters of the structure and the main characteristics of the coupled modes, such as group velocity and local and average polarization of electric field, we have applied a coupled mode theory (CMT)~\cite{YarivBook,SukhorukovJOSAB2008}. Throughout the text we numerate modes in such a way that indices ``1'' and ``2'' correspond to the modes of the periodic waveguide and index ``3'' -- to the mode of the homogeneous waveguide. We assume, that the field distribution of a mode of the coupled waveguides $\VB{E}$ can be represented as a superposition of the modes of the isolated waveguides $\VB{E}_i$, $i=1,2,3$ in the following way:
\begin{equation}
\VB{E}(\VB{r},k) = \sum_{i}a_i(k)\VB{E}_i(\VB{r},k),
\end{equation}
where $\VB{r}$ is the radius vector, $k$ is the Bloch wavenumber, and $a_i(k)$ is the complex amplitude of the $i$th mode. Then, the eigenfrequencies $\omega$ of the coupled modes as a function of $k$ can be found from the following set of equations:
\begin{equation}
\label{eq:CMT_0}
\VB{M}(k,\omega)\VB{A}(k) = 0,
\end{equation}
where $\VB{A} = (a_1(k),a_2(k),a_3(k))^T$, and the elements of the coupling matrix $\VB{M}(k)$ are defined via the overlap integrals:
\begin{eqnarray}
\label{eq:M_full}
M_{ij}(k,\omega) = (\omega^2 - \omega_j^2(k))A_{ij}(k) + \omega^2B_{ij}(k),\\
\label{eq:diag_integr}
A_{ij}(k) = \langle \varepsilon_j(\VB{r}) \VB{E}_i^*(\VB{r},k)\cdot \VB{E}_j(\VB{r},k) \rangle,\\
\label{eq:non_diag_integr}
B_{ij}(k) = \langle (\varepsilon(\VB{r}) - \varepsilon_j(\VB{r}))\VB{E}_i^*(\VB{r},k)\cdot \VB{E}_j(\VB{r},k) \rangle,
\end{eqnarray}
where $i,j=1,2,3$, $\omega_j(k)$ defines the dispersion of the $j$th mode, averaging is performed over the unit cell $\langle \ldots \rangle = \int_{0}^{a} \int_{-\infty}^{+\infty} \int_{-\infty}^{+\infty}  \ldots dx dy dz$, $a$ is the period of the structure in $z$ direction, $\varepsilon({\bf r})$ is the permittivity of the coupled waveguides and $\varepsilon_j({\bf r})$ is the permittivity of the isolated waveguide supporting $j$th mode. 

When the periodic and homogeneous waveguide are brought together, the coupling between the modes of the periodic waveguide ($M_{12}$) occurs only via the homogeneous waveguide perturbation. In the considered structures it is much weaker than the coupling between the modes of the periodic waveguide and the mode of a homogeneous waveguide ($M_{13}$ or $M_{23}$). However, as it is shown further, to obtain the desired characteristics of the guided structure we need to have comparable strength of coupling between all of the modes. To achieve that, we introduce a local perturbation in the periodic waveguide (schematics are shown in Sec. C), which almost does not affect the elements $M_{13},M_{23},M_{31},M_{32}$, while the elements $M_{12}$ and $M_{21}$ become of the same order of magnitude as the other non-diagonal elements. In order to correctly find the coupling between the modes ``1'' and ``2'' caused by this perturbation, one need to modify the integral~\eqref{eq:non_diag_integr} as follows~\cite{JohnsonPRE2002}:
\begin{equation}
\label{eq:non_diag_integr_corr}
\tilde{B}_{ij}(k) = \langle (\varepsilon(\VB{r}) - \varepsilon_j(\VB{r}))\VB{E}_{i\|}^*(\VB{r},k)\cdot \VB{E}_{j\|}(\VB{r},k) - (1/\varepsilon(\VB{r}) - 1/ \varepsilon_j(\VB{r}))D_{i\perp}^*(\VB{r},k)\cdot D_{j\perp}(\VB{r},k) \rangle,
\end{equation}
where $\VB{E}_{\|}$ is the vector $\VB{E}$ with zero component normal to the perturbation interface, and $D_{i\perp}$ is the component of the electric displacement field vector normal to the perturbation interface. After estimating the order of magnitude of all elements in~\eqref{eq:M_full} and neglecting all terms larger than the first order, we rewrite the matrix equation~\eqref{eq:CMT_0} as
\begin{equation}
\label{eq:CMT}
{\bf M}(k){\VB{A}(k)} =\omega(k) {\VB{A}(k)},
\end{equation}
where the elements of the newly defined coupling matrix are simplified to:
\begin{equation}
\begin{aligned}
\label{eq:CMT_main}
M_{ii}(k) = \omega_i(k),\\
M_{ij}(k) = - \dfrac{\omega_j^2(k)}{\omega_i(k)}\dfrac{\tilde{B}_{ij}(k)}{2A_{ii}(k)},\; ij=\{12,21\},\\
M_{ij}(k) = - \dfrac{\omega_j^2(k)}{\omega_i(k)}\dfrac{B_{ij}(k)}{2A_{ii}(k)},\; ij=\{13,23,31,32\}.
\end{aligned}
\end{equation}
When calculating $M_{12}$ and $M_{21}$ elements, integration~\eqref{eq:non_diag_integr_corr} is performed only over the perturbation introduced in the periodic waveguide, while for $M_{13},M_{23},M_{31},M_{32}$ elements this perturbation can be neglected in the integrals~\eqref{eq:non_diag_integr}.

To derive the requirements to the elements of the coupling matrix that allow to achieve the dispersion with a stationary inflection point (SIP) and circular polarization of the local electric field in the periodic waveguide, we impose the following conditions on the eigenmode of the Eq.~\eqref{eq:CMT} for a wavenumber $k_0$:
\begin{eqnarray}
    \label{eq:SIP_condition}    \dfrac{d\omega}{dk}\Bigr|_{k=k_0}=0,\;\;\dfrac{d^2\omega}{dk^2}\Bigr|_{k=k_0}=0,\\
    \label{eq:CP_condition}
    \VB{A}(k_0) = (1,i\alpha,\beta)^T,
\end{eqnarray}
where $\alpha$ and $\beta$ are arbitrary real numbers (the phase of the third mode can be chosen at will). Note that $k_0$ is not a wavenumber at the crossing point of the modes ``1'' and ``2'' but rather an arbitrary wavenumber. The Eq.~\eqref{eq:SIP_condition} specifies the existence of the SIP, $\omega(k) \approx \omega(k_0) + \gamma(\delta k)^3$, at the point $k_0$, while the eigenvector~\eqref{eq:CP_condition} ensures the desired polarization of the field in the periodic waveguide. The phase difference between the components is fixed as $\pm \pi/2$, while the amplitude ratio $|\alpha|$ depends on the field distribution of the first and second modes and the choice of the normalization. For instance, $|\alpha|$ can be chosen in such a way that the amplitudes of the local electric field of two constituent modes at a certain point are equal to each other, thus realizing local circularly polarized electric field.

The relation between the elements of the coupling matrix that satisfy the condition~\eqref{eq:CP_condition} can be found by straight forward substitution of the required eigenvector at a fixed value of $k_0$ into Eq.~\eqref{eq:CMT}. The conditions~\eqref{eq:SIP_condition}, on the other hand, require the knowledge of the decomposition of the matrix elements up to second order over the wavenumber detuning $\delta k = k - k_0$:
\begin{equation}
\label{eq:decomposition}
\begin{aligned}
\omega_i(k) = \omega_i(k_0) + v_{gi}\delta k + D_{i}\delta k^2,\\
|M_{ij}|^2 = |M_{ij}|^2_0 + |M_{ij}|^2_1\delta k + |M_{ij}|^2_2\delta k^2,\; i \ne j,
\end{aligned}
\end{equation}
where $v_{gi}$ and $2D_i$ are the group velocity and group velocity dispersion of the $i$th mode, respectively. The large amount of the variables emerging after such decomposition significantly complicates the analysis and makes it practically useless. However, in realistic structure one can more or less control only the zeroth order terms, i.e. the frequency shift of the isolated modes and the strength of coupling between the modes, but not the first and second order terms of the decomposition. Further numerical estimations performed for the realistic designs have shown that the qualitative as well as quantitative requirements to the parameters of the system can be identified with a good level of approximation even after neglecting some of the elements. Namely, the matrix $\VB{M}$ can be simplified to:
\begin{equation}
\begin{aligned}
M_{ii}(\delta k) = \omega_i(k_0) + v_{gi}\delta k,\\
M_{ij} = - \dfrac{\omega_j^2(k_0)}{\omega_i(k_0)}\dfrac{B_{ij}(k_0)}{2A_{ii}(k_0)},\; i \ne j,
\end{aligned}
\end{equation}
where $v_{gi}$ is the group velocity of the $i$th mode.
By introducing the frequency detuning $\widetilde{\omega}(k) = \omega(k) - \widetilde{\omega}_0$, where $\widetilde{\omega}_0 = \widetilde{\omega}_1 (k_0) = \widetilde{\omega}_2 (k_0)$, and normalizing the fields in such a way that $M_{ij} = M_{ji}^*$, we rewrite the Eq.~\eqref{eq:CMT} in the following form: 
\begin{equation}
\label{eq:CMT_simple}
\begin{pmatrix} v_{g1}\delta k & M_{12} & M_{13} \\
M_{12}^* & v_{g2}\delta k & M_{23} \\
M_{13}^* & M_{23}^* & \Delta + v_{g3}\delta k
\end{pmatrix} {\VB{A}(k)} = \widetilde{\omega}(k) {\VB{A}(k)},
\end{equation}
where $\Delta = \widetilde{\omega}_3 (k_0) - \widetilde{\omega}_0$. The simplified form of the CMT~\eqref{eq:CMT_simple} is mainly used in the further analysis.

\subsection{Application of the CMT to the first design}

First, we note that the quasi-degenerate modes of the periodic waveguide have orthogonal $y$ and $z$ dominant components of the fields, while the evanescent field of the third mode in the region, where periodic waveguide is placed, has both $y$ and $z$ components with comparable amplitudes and with $\pm \pi/2$ phase difference. Consequently, the elements of the coupling matrix $M_{13}$ and $M_{23}$ are expected to have a fixed phase difference $\pm \pi/2$. The absolute values of these constants also depend on the field profiles of the modes ``1'' and ``2''. More specifically, on the amplitude of the term with $n=-1$ in the Fourier decomposition, according to the phase matching condition. The ratio $M_{13}/M_{23}$ extracted from the numerical simulations was equal to -0.85i. Taking this into account, one can notice that there is a simple solution that satisfies the conditions~(\ref{eq:SIP_condition}-\ref{eq:CP_condition}) when all three modes intersect at the same point $k_0=0$, i.e. when $\Delta=0$. Then, for a zero wavenumber detuning $\delta k$ the system has an eigenvector:
\begin{equation}\label{eq:CP}
\VB{A}=(1,M_{13}/M_{23},-M_{12}/M_{23})^T,
\end{equation}
with the eigenfrequency $\omega=0$, if the coupling constants $M_{12}, M_{23}$ are real and $M_{13}$ is imaginary. Furthermore, the condition
\begin{equation} \label{eq:ZGV}
v_{g1}|M_{23}|^2 + v_{g2}|M_{13}|^2 + v_{g3}|M_{12}|^2 = 0.
\end{equation}
ensures that the group velocity of the coupled mode vanishes, while the group velocity dispersion vanishes automatically.

The condition~\eqref{eq:ZGV} can be satisfied by choosing an appropriate ratio of the amplitudes of the $M_{13}$, $M_{23}$ and $M_{12}$ elements. The Eq.~\eqref{eq:CP} shows that the amplitude of the second mode $\alpha$ is equal to the ratio $M_{13}/M_{23}$, which is close to the desired value $i$ in the realistic systems. Since all group velocities are of the same order and $v_{g1}$ and $v_{g2}$ have the same sign, the matrix element $M_{12}$ has to be of the same order as $M_{13}$ or $M_{23}$. This can be achieved by introducing the deformation in the periodic waveguide that couples its modes strongly enough. The symmetry of the deformation that provides the real valued $M_{12}$ is discussed in the Sec.~C.

The parameters and characteristics of the considered structure (Fig.~3(a) in the main text) that were determined according to the above-mentioned recipe, turned out to be slightly different than those expected from the CMT. Further analysis revealed that such discrepancy is caused by the non-negligible coupling of the modes of the periodic waveguide with another mode of the homogeneous waveguide (lower branch of the third mode in Fig.~\ref{fig:folding}). Although this mode is well separated in frequency, the interaction with it can be stronger than that with the backward branch due to direct coupling (not contradirectional coupling via reciprocal lattice vector). Although in order to find the characteristics of the system with a very fine level of accuracy one need to include this mode into consideration, the general concept remains the same and with high level of accuracy even when considering the coupling only between three main modes.

Note also, that if one takes into account all first and second order terms in the decomposition of the matrix elements over $\delta k$ in Eqs.~\eqref{eq:decomposition}, the group velocity dispersion of the middle branch at $k_0$ is modified in the following way:
\begin{equation}
\label{eq:GVD_full}
D = \dfrac{D_1 + D_2|\alpha|^2 + D_3|\beta|^2 + v_{g1}\dfrac{|M_{23}|^2_1}{|M_{23}|^2_0}+ v_{g2}\dfrac{|M_{13}|^2_1}{|M_{23}|^2_0} + v_{g3}\dfrac{|M_{12}|^2_1}{|M_{23}|^2_0}}{1 + |\alpha|^2 + |\beta|^2}.
\end{equation}
For the wavenumber $k_0=0$ group velocity $v_g$ is still equals to zero, but according to~\eqref{eq:GVD_full} $D$ is non-zero, which breaks the condition of the SIP existence. However, SIP can be restored by a very slight change of the $k_0$ and e.g. $M_{12}$. In realistic structures the values of $D_i$ and $|M_{ij}|^2_{1,2}$ are so small, that the required changes in $k_0$ and $M_{12}$ are much less than 1\%, therefore, the effect of higher order terms in the decomposition~\eqref{eq:decomposition} can be neglected.

\subsection{Application of the CMT to the second design}

In the second design we assume that the second mode of the periodic waveguide has negative velocity, while the first mode of the periodic waveguide and the mode of the homogeneous waveguide (third mode) have positive group velocities. Because the coupling between the modes ``1'' and ``3''  is codirectional, and between the modes ``2'' and ``3'' it is contradirectional, typically the coupling constant $M_{23}$ is much less than $M_{13}$. Therefore, for the second design for simplicity we have neglected the coupling $M_{23}$ and considered the coupling matrix in the following form:
\begin{equation}
\VB{M} = \begin{pmatrix} v_{g1}k & M_{12} & M_{13} \\
M_{12}^* & v_{g2}k & 0 \\
M_{13}^* & 0 & v_{g3}k + \Delta
\end{pmatrix}.
\end{equation}
Applying the conditions~(\ref{eq:SIP_condition}-\ref{eq:CP_condition}) provides us with the relation between the parameters of the system:
\begin{equation}
\begin{aligned}
\label{eq:Solution_2des}
& M_{12} = -i\dfrac{k_0 \alpha^3(v_1-v_2) v_2^2}{(v_1^2+2\alpha^2 v_1 v_2 + \alpha^2 v_2^2)}, \\
& M_{13} = \pm \dfrac{k_0 (v_1-v_2)\sqrt{-(v_1+\alpha^2 v_2)^3 v_3}}{(v_1^2+2\alpha^2 v_1 v_2 + \alpha^2 v_2^2)}, \\
& \Delta = \dfrac{k_0 v_1 [v_1(v_2-2v_3) + v_2(3\alpha^2 (v_2-v_3)+v_3)]}{(v_1^2+2\alpha^2 v_1 v_2 + \alpha^2 v_2^2)}, \\
& \omega_0 = \dfrac{k_0 v_1 v_2 (v_1+3\alpha^2 v_2)}{(v_1^2+2\alpha^2 v_1 v_2 + \alpha^2 v_2^2)}, \\
& \beta = - \sign(M_{13}) {\sqrt{-\dfrac{v_1+\alpha^2 v_2}{v_3}}}.
\end{aligned}
\end{equation}

Relations~\eqref{eq:Solution_2des} show that a SIP with a desired polarization can be realized for an arbitrary value of $k$, except the trivial case of $k=0$. Since $M_{12}$, $M_{13}$ are proportional to $k_0$, while they should be relatively small, we also aim at small values of $k_0$. According to~\eqref{eq:Solution_2des} the constant $M_{12}$ in the second design should be imaginary. This can be achieved e.g. by shifting the holes in transverse direction, see Sec.~C for details. Note, that the shift of the holes is accompanied by the frequency shift of the modes, i.e. by appearance of the non-zero diagonal elements $M_{11}$ and $M_{22}$. Such modification, however, only requires to measure the frequency and the wavenumber detuning from the new crossing point of the shifted modes ``1'' and ``2''.

\begin{figure*}[t]
    \includegraphics[width=1\textwidth]{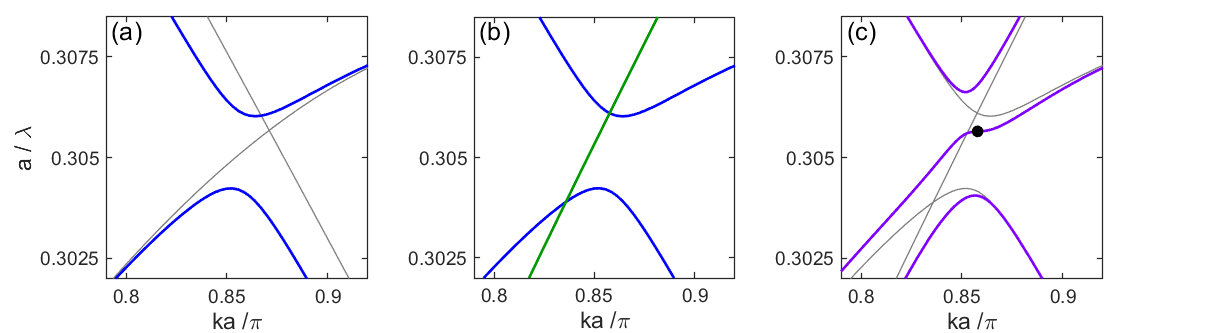}
    \caption{(a) Dispersion of the modes of a periodic waveguide with symmetrically placed holes (gray curves) and shifted holes (blue curves). (b) Blue curves -- the same as in panel (a), green curve shows the dispersion of the mode of an isolated homogeneous waveguide. (c) Gray curves -- the same as blue and green curves in panel (b), purple curves show the dispersion of the coupled periodic and homogenous waveguides. Black dot indicates a SIP.} 
    \label{fig:APP_design2}
\end{figure*}

One can also understand the dispersion, shown in Fig.~2(b) in the main text and obtained with the use of the conditions~\eqref{eq:Solution_2des}, in a different way, by considering the coupling between the modes in two steps. At the first step one we consider only the coupling between the modes of the periodic waveguide. Introducing perturbation to the waveguide couples two modes with the phase difference dictated by the coupling constant $M_{12}$. The dispersion of the modes of the unperturbed and perturbed periodic waveguide are shown in Fig.~\ref{fig:APP_design2}(a) with thin gray and thick blue curves, respectively. As was mentioned earlier, due to the non-zero diagonal terms of the coupling matrix along with the formation of the anticrossing there is a slight shift of the dispersion. By choosing the imaginary constant $M_{12}$, upper and lower dispersion curves acquire elliptical polarization with either $\pi/2$ or $-\pi/2$ phase difference between the constituent modes. At the second step, we choose the frequency shift of the third mode $\Delta \approx M_{12}$, as shown with the green line in Fig.~\ref{fig:APP_design2}(b). Then, the third mode couples to the upper branch of the perturbed periodic waveguide at the point with negative group velocity. Consequently, by choosing the appropriate value of the $M_{13}$ constant we obtain the SIP with the polarization defined mainly by the constant $M_{12}$ at the first step, see purple curves in Fig.~\ref{fig:APP_design2}(c).

\section*{C. Relation between the geometrical parameters of the nanostructure and coupling constants in the CMT}\label{sec:AppC}

Although all variables in the CMT depend on the geometrical parameters of the structure, the main parameters that can be controllably tuned by the geometrical adjustment are the frequency shifts of the modes of isolated waveguides and coupling constants at a given wavenumber. 

Since the operational frequency is always close to the crossing point of the dispersion curves of two modes of the periodic waveguide, we measure the frequency shift $\Delta$ of the third mode from this frequency. This is done straight forwardly by changing the width of the strip waveguide $w$. Decrease (increase) of the waveguide thickness results in the increase (decrease) of the frequency of the waveguide mode for a given wavenumber. For small values of the detuning, relevant to the current designs, $\Delta $ is linearly proportional to the $\delta w$, thus allowing for simple tuning of the diagonal element $M_{33}$ of the coupling matrix.

The elements $M_{13}$ and $M_{23}$ describe the coupling between the modes of the periodic waveguide and the mode of the homogeneous waveguide. The phase difference between these elements is fixed as $\pi/2$, while their absolute values are determined by two factors. First, the less is the separation between the waveguides, the larger are the coupling constants. Second, since in some cases the coupling between the modes is contradirectional, the efficiency of coupling also depends on the $n=-1$ component in the Fourier decomposition of the field profiles of the periodic waveguide modes. Typically, a shift of the operational wavenumber closer to the edge of the Brillouin zone increases the amplitude of the minus first Fourier component of the mode and thus increases the coupling strength.

\begin{figure*}[t]
    \includegraphics[width=0.6\textwidth]{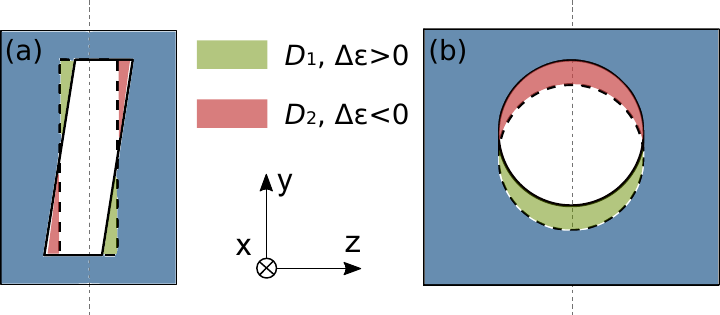}
    \caption{Schematic illustration of the two types of asymmetric perturbation in the unit cell of a periodic waveguide. Dashed curve shows the boundary of a hole in the unperturbed waveguide, solid curve shows the modified position of the hole. (a) Perturbation is symmetric with respect to rotation around the $x$ axis by 180 degrees. (b) Symmetry with respect to the $z=0$ plane is preserved. } 
    \label{fig:APP_asymmetry}
\end{figure*}

The element $M_{12}$ can be controlled by introducing a local perturbation in the periodic waveguide. The value of $M_{12}$ is proportional to the overlap integral $\int_{\text{unit cell}}  \Delta\varepsilon(\VB{r})C(\VB{r})  d\VB{r}$,~\eqref{eq:non_diag_integr_corr}, where $\Delta\varepsilon$ describes the change of the permittivity as compared to the unperturbed waveguide. Taking into account that $y=0$ is magnetic (electric) symmetry plane of the first (second) mode, and that rotation around the $x$ axis transforms the mode into its time reversal partner with complex conjugate field, one gets the following properties of the function $C$: $C(-y) = -C(y)$, $C(-z) = -C^*(z)$. Therefore, non-zero coupling is achieved by the perturbation that breaks the symmetry with respect to the $y=0$ plane. In general, the coupling constant $M_{12}$ is complex-valued. However, if the symmetry of the perturbation $\Delta\varepsilon$ with respect to $z=0$ plane is preserved, the element $M_{12}$ becomes purely imaginary. And if instead the perturbation has a rotation symmetry around $x$ axis, the coupling is purely real. This is illustrated in Fig.~\ref{fig:APP_asymmetry}.

Besides the coupling between the modes, such perturbation also induces the spectral shift of the modes, i.e. non-zero diagonal elements $M_{11}$ and $M_{22}$. These can be partially compensated if along with the positive change of the permittivity in some region (dielectric instead of vacuum, $\Delta \varepsilon>0$), there is also negative change in another region (vacuum instead of dielectric $\Delta \varepsilon <0$). For instance, in Fig.~\ref{fig:APP_asymmetry}(a), the domains $D_1$ provide negative contribution to the overlap integral, while $D_2$ -- positive contribution. Note, that such compensation is not ideal, because of the slightly different field distributions in the domains $D_1$ and $D_2$ in the unperturbed waveguide.

\bibliography{References}